\documentclass[12pt]{article}
\usepackage{epsfig}
\usepackage{psfrag}
\def\beq{\begin{equation}}
\def\eeq{\end{equation}}
\def\bea{\begin{eqnarray}}
\def\eea{\end{eqnarray}}
\def\bq{\begin{quote}}
\def\eq{\end{quote}}

\def\AJ{{\it Astrophys. J.} }

\def\AP{{\it Ann. Phys.} }

\def\CQG{{\it Class. Quantum Gravity} }

\def\GRG{{\it Gen. Relativity and Gravitation} }

\def\IJMP{{\it Int. J. Mod. Phys.} }

\def\NC{{\it Nuovo Cimento} }
\def\NP{{\it Nucl. Phys.} }
\def\PL{{\it Phys. Lett.} }
\def\PR{{\it Phys. Rev.} }
\def\PRL{{\it Phys. Rev. Lett.} }
\def\PRTS{{\it Physics Reports} }

\def\vev#1{\langle{#1}\rangle}

\def\lsim{\mathrel{\rlap{\lower4pt\hbox{\hskip1pt$\sim$}}
    \raise1pt\hbox{$<$}}}
\def\gsim{\mathrel{\rlap{\lower4pt\hbox{\hskip1pt$\sim$}}
    \raise1pt\hbox{$>$}}}
\def\sqr#1#2{{\vcenter{\vbox{\hrule height.#2pt
         \hbox{\vrule width.#2pt height#1pt \kern#1pt
         \vrule width.#2pt}
         \hrule height.#2pt}}}}

\def\gappeq{\mathrel{\rlap {\raise.5ex\hbox{$>$}}
{\lower.5ex\hbox{$\sim$}}}}

\def\lappeq{\mathrel{\rlap{\raise.5ex\hbox{$<$}}
{\lower.5ex\hbox{$\sim$}}}}

\begin{document}
\pagestyle{empty}

\begin{flushright}
{\sc DF/IST-4.98} \\
{\sc FEBRUARY 1999} \\
\end{flushright}

\newcommand\figinsert[4]
{\begin{figure}[htb]
\newcommand\figsize{#3}
\epsfysize\figsize
\vskip -.5cm
\centerline{\epsffile{#4}}
\vskip -.0cm
\caption{\leftskip 1pc \rightskip 1pc
\baselineskip=10pt plus 2pt minus 0pt {{#2}}}
\label{fig #1}
\vskip -.3cm\end{figure}}

\setlength{\baselineskip}{0.75cm}

\vspace*{1.0cm}

\begin{center}
{\large\bf Nucleosynthesis Constraints on a 
Scale-Dependent New Intermediate Range Interaction}\\

\vspace*{2.0cm}
{\bf O. Bertolami$^{1)}$  and F.M.  Nunes$^{2)}$}\\
\medskip
{1) Depto. de F\'\i sica, Instituto Superior T\'ecnico, 1096 Lisboa Codex, 
Portugal}\\
{2) Universidade Fernando Pessoa, Pra\c{c}a 9 de Abril, 4200 Porto, Portugal}\\

\vspace*{2.0cm}
{\bf ABSTRACT} \\ 
\end{center}

\noindent
We derive constraints on the strength 
of a new intermediate range interaction that couples to baryon number
from primordial nucleosynthesis yields.
The nucleosysnthesis limits here used
arise from matching observations and predictions of standard
and inhomogeneous primordial scenarios.
We show that the standard nucleosynthesis scenario is 
more restrictive ($\alpha_5 \lsim 0.2$)  when the range of the interaction is 
greater than about
$1~m$. We further discuss the implications of considering the scalar 
particle responsible for the new interaction as the main component
of the dark matter in the galactic halo such that its decay can account
for the ionization of hydrogen in the interstellar medium and the 
temperature of Lyman-$\alpha$ clouds.

\vfill
\eject

\setcounter{page}{2}
\pagestyle{plain}

\vspace{0.5cm}

1. The existence of new fundamental forces beyond the already known four
interactions is an exciting possibility that may have profound implications 
in our understanding of the physics beyond the standard model. The 
claim, more than a decade ago, of evidence for an intermediate range
interaction with sub-gravitational strength \cite{fishbach1} has led to
a great demand of theoretical explanations (see \cite{nieto} for a review
and a complete set of references) and, most importantly, has given
origin to fresh experiments based on new ideas and to 
the repetition of well known experiments using new state of the art technology.
  
In its simplest versions, the putative new interaction or a {\it 
fifth force} would arise from
the exchange of a light boson coupled to matter with a strength
comparable to gravity. There are several schemes through which physics 
at the Planck scale could give origin to such an interaction and yield 
a Yukawa type modification in the interaction energy between 
point masses. This new contribution to the
interaction energy can arise, for instance, from
extended supergravity theories after dimensional reduction 
\cite{nieto, scherk}, from the compactification of $5$-dimensional
generalized Kaluza-Klein theories that include gauge interactions at
higher dimensions \cite{bars}  and also from string theory. 
On quite general terms, the interaction energy, $V(r)$, between two 
point masses $m_1$ and $m_2$,
can be expressed in terms of the gravitational interaction as

\begin{equation}
V(r) = -  \frac{G_{\infty}~m_{1}~m_{2}}{r}~(1 + \alpha_5~e^{-r/\lambda_5})~,
\label{eq:1.1}
\end{equation}
where $r = \vert \vec r_{2} - \vec r_{1} \vert$ is the distance between the
masses, $G_{\infty}$ is the gravitational coupling for $r \rightarrow \infty$, 
$\alpha_5$ and $\lambda_5$ are the strength and the range of the new 
interaction.
Of course, $G_{\infty}$ has to be identified with the Newtonian gravitational 
constant and the gravitational coupling would be dependent on $r$.
Indeed, the force associated with eq. (\ref{eq:1.1}) is given by:

\begin{equation}
\vec F(r) = -  \nabla V(r) = - \frac{G(r)~m_{1}~m_{2}}{r^2}~\hat {\bf r}~,
\label{eq:1.2}
\end{equation}
where

\begin{equation}
G(r) = G_{\infty}[1 + \alpha_5~(1 + r/\lambda_5)~e^{-r/\lambda_5}]~.
\label{eq:1.3}
\end{equation}

The great interest sparked by the suggestion of existence of
a new interaction 
was the recognition that the coupling $\alpha_5$ was not an universal 
constant, but instead a parameter depending on the chemical composition
of the test masses \cite{lee}. This dependence comes about if one assumes 
that the  new
bosonic field couples to the baryon number $B = Z + N$ which is the sum
of protons and neutrons. Hence the new interaction between masses with 
baryon numbers $B_1$ and $B_2$ can be expressed through a new
fundamental constant, $f$, in the following way:

\begin{equation}
V(r) = - f^{2}~\frac{B_{1}~B_{2}}{r}~e^{-r/\lambda_5}~,
\label{eq:1.4}
\end{equation}
such that the constant $\alpha_5$ can be written as
\begin{equation}
\alpha_{5}= - \sigma~\left(\frac{B_{1}}{\mu_{1}}\right)~
\left(\frac{B_{2}}{\mu_{2}}\right)~,
\label{eq:1.5}
\end{equation}
with $\sigma = f^{2}/G_{\infty}m_{H}^{2}$ and $\mu_{1,2} = m_{1,2}/m_{H}$
($m_H$ is the hydrogen mass). 

Of course, from the above equations it follows that in a Galileo-type
experiment an acceleration difference between masses $m_1$ and $m_2$
would be given by:
 
\begin{equation}
\vec a_{12}= \sigma~\left(\frac{B}{\mu}\right)_{\oplus}
\left[\left(\frac{B_{1}}{\mu_{1}}\right)~-~
\left(\frac{B_{2}}{\mu_{2}}\right)\right]~\vec F~,
\label{eq:1.6}
\end{equation}
where $\vec F$ is the field strength of the Earth (which is denoted by 
$\oplus$). 

Several experiments (see, for instance, \cite{fishbach1} for a list of
the most important ones) have been performed in order to establish the 
parameters of a new interaction based on the idea of a composition-dependence 
differential acceleration as described in eq. (\ref{eq:1.6}) and other
composition-independent effects. The current experimental
situation is essentially compatible with predictions of Newtonian gravity
in either composition-independent or composition-dependent experiments. The  
bounds on parameters $\alpha_5 $ and $\lambda_5$ can be summarised as follows:\\
 - satellite tests probing ranges about $10^{5}~m < \lambda_5 < 10^{7}~m$ 
indicate that $\alpha_5 < 10^{-5}$; \\
 - gravimetric experiments that are sensitive in the range of
$10~m < \lambda_5 < 10^{3}~m$ suggest that $\alpha_5 < 10^{-3}$;\\
 - laboratory experiments deviced to measure deviations from 
the inverse-square law are sensitive essentially to the range 
$10^{-2}~m < \lambda_5 < 1~m$ and constrain $\alpha_5$ to be smaller than 
$10^{-4}$.

Interestingly, for $\lambda_5 < 10^{-3}~m$ and  
$\lambda_5 > 10^{13}~m$, $\alpha_5$ is essentially unconstrained. 
The former range is particularly attractive as new forces with 
sub-millimetric range seems to be favoured from scalar interactions in 
supersymmetric theories \cite{dimopoulos} and in the recently proposed 
theories on $TeV$ scale quantum gravity \cite{arkani}. 
This range also arises from 
assuming that scalar \cite{beane} or tensor interactions associated to
the breaking of the Lorentz invariance in string theories \cite{bertolami1}
account for the vacuum energy up to the level $\Omega_{V} < 0.5$.
  
In order to close our summary of the experimental situation we should 
point out that, as discussed in \cite{fishbach1}, existing experimental 
data cannot account for certain anomalies such as the one claimed to exist 
in the original E\"otv\"os experiment \cite{fishbach2} and hence 
these anomalies remain still an open issue. Another quite recent claim
concerning the existence of a new interaction, is given by the radio metric data
from Pioneer 10/11, Galileo and Ulysses spacecraft, indicating an anomalous
constant acceleration of about $8.5 \times 10^{-8}~cm~s^{-2}$ acting
on the spacecraft directed towards the sun. A new interaction with  
$\alpha_5 = - 1 \times 10^{-3}$ and 
$\lambda_5 = 200~A.U. = 1.49 \times 10^{13}~m$ seems to account for the
anomaly \cite{anderson} (see however \cite{katz}).

In this work we shall establish limits to the coupling
$\alpha_5$ and range $\lambda_5$ of the putative new interaction 
using results from nucleosynthesis in the context of the Big-Bang. 
As we shall see, these limits appear to be much less stringent than the ones 
obtained from laboratory experiments. Nevertheless, independently
from the theoretical setting from which 
this interaction may arise, if the coupling of the putative new interaction
is a running coupling constant, the limits derived here may turn out
to be relevant.  
Indeed, in this case, 
the coupling constant at the time of primordial nucleosynthesis 
$\alpha_5^{prim}$ 
can be greater than the bounds arising 
from laboratory experiments, as long as the interaction is a local 
gauge interaction and its $\beta$-function is positive. 
This is, for instance,
the case of the $U(1)$ coupling constant in QED.
Assuming scale-dependence for the new interaction coupling 
constant is quite a natural 
assumption, as all gauge coupling constants in the standard model are running.
Even the gravitational coupling constants are, 
at one-loop level, running couplings in higher derivative
theories of gravity \cite{julve} (including of course the Newton's constant).
This fact has implications for quite a few diversified problems:
the problem of the rotation curve of galaxies 
\cite{bertolami2,bottino}; the cosmological dark matter
problem \cite{goldman}, the large scale structure of the Universe 
\cite{bertolami3} and the cosmic virial theorem \cite{bertolami4}.
Notice that we are concerned only with zero temperature running 
effects as those for couplings of the type $g \phi \psi \bar \psi$ 
(cf. eq. (\ref{eq:3.1})). The finite temperature corrections that are
proportional to $g^2$ (see eg. \cite{chia}) become negligible 
(cf. considerations after eq. (\ref{eq:3.2})) in the limit $T >> m_{5}$ 
(since we shall deal with temperatures $T \sim MeV$ and argue that 
$m_{5} \sim eV$).

Of course, the limits that we are going to establish can be regarded
as independent of any considerations concerning the running of couplings 
and are on their own of relevance as they are consistent with laboratory
experiments and are obtained from an independent line of reasoning than
the usual approaches.

Standard primordial nucleosynthesis scenario (see \cite{olive} for a review)  
allows obtaining bounds on the
variation of the effective gravitational coupling when 
confronted with the measurements of abundance
of light elements in the Universe. Such measurements are in 
agreement with the standard nucleosynthesis scenario
leaving still some room for 
variations in the effective number of neutrinos, the baryon fraction 
of the universe and also in the value of the gravitational constant.
It is well known that the predicted mass fraction of primordial $^4$He
can be parametrised, in theories with a variable gravitational coupling,
in the following way \cite{olive,casas},

\begin{equation}
Y_{\rm p} = 0.228 + 0.010 \ln \eta_{10} + 0.327 \log \xi\ ,
\label{eq:1.7}
\end{equation}
where $\eta_{10}$ denotes the baryon to photon ratio in units of $10^{-10}$
and $\xi$ is the ratio of the Hubble parameter at the time of  nucleosynthesis 
and its present value 
which is itself proportional to the square root of the
gravitational constant. In the fit eq. (\ref{eq:1.7}) it is
assumed that the effective number of light neutrinos is $N_\nu = 3$
and that the neutron lifetime is $\tau_n = 887.0 \pm 2.0$ seconds \cite{PDG}.

A range for the values of the effective gravitational coupling that are
compatible with the observations of the primordial D, $^3$He, 
$^4$He and $^7$Li abundances can be obtained
running the nucleosynthesis codes for different values of $G$
\cite{accetta} . It turns out that the permissible range is rather large
($\Delta G/G = 0.2$ at the $1~\sigma$ level), given the large statistical 
and systematic errors of the observations.
We shall use this result  to constrain the parameters of a new intermediate
range interaction assuming it to be sensitive to baryon number. The 
maximum range that can be tested corresponds to the physical
horizon distance at the time of primordial nucleosynthesis.  At that time, 
the horizon distance grows from a few light-seconds to 
a few light-minutes,  and therefore is much smaller than a few microparsecs:
$r < 3 \times 10^{10}~m$. Using eq. (\ref{eq:1.3}) and considering
the  $1~\sigma$ level result of Ref. \cite{accetta}, we obtain:

\begin{equation}
\frac{\Delta G(r)}{G_{\infty}} = \alpha_5~(1 + r/\lambda_5)~e^{-r/\lambda_5} < 
0.2~~,
\label{eq:1.8}
\end{equation}
which we shall use in the next sections to impose constraints on the
parameters of the new interaction.

Of course, as a light-second is about the distance to the Moon, 
constraints on a variation of $G$ can, at this scale, be set 
from lunar laser ranging, $\Delta G/G < 0.6$ \cite{will}. 
It is worth mentioning  that similar reasoning has 
been used to impose constraints on the
variation of the gravitational coupling due to a possible dependence on scale
\cite{bertolami2}. 

%%%%%%%%%%%%%%%%%%%%%%%%%%%%%%%%%%%%%%%%%%%%%%%%%%

2. Let us turn to the discussion of the  bounds that can be imposed from
the standard primordial nucleosynthesis scenario.  First of all we need to 
estimate
a typical distance $r$ between particles interacting through this new 
interaction. 
Only then can we infer on the range and strength of the fifth force. 

The period of interest is that immediately 
after the weak-interaction decoupling,  for which temperatures are of the 
order  $E\sim 1$ MeV
and the total baryon density is  
$\rho_B  \sim 1.21 \times 10^{-2}~g~cm^{-3}$ \cite{wagoner}. 
We shall assume that, in this epoch, the baryons 
are in the form of protons and neutrons only: $\rho_B = \rho_p + \rho_n$.

One can argue that the typical distance between the interacting 
particles should be smaller or of the same
order as their mean free paths $\lambda_n$ and $\lambda_p$. 
In order to estimate $\lambda_n$ and $\lambda_p$ we need 
to calculate the relevant 
stopping cross sections and the densities of the stopping particles.
Within a standard scenario the number of neutrons is slightly smaller
than the number of protons due to the mass difference and neutron decay:
\begin{equation}
\frac{n_n}{n_p} = e^{-\frac{\Delta m \: c^2}{kT} } \: e^{- \frac{t}{\tau_{n} 
}} \;\;
\simeq \; \; \frac{1}{e}\;,
\label{eq:2.1}
\end{equation}
although, 
at this temperature $t << \tau_{n}$, and hence the neutron decay is not
so important.  
Using eq. (\ref{eq:2.1}) and the charge neutrality condition, we arrive
at the following estimates:
$\rho_p = 0.9 \times 10^{-2}~g~cm^{-3}$, $\rho_n = 0.3 \times 10^{-2}~g~ 
cm^{-3}$, and $\rho_e = 0.5 \times 10^{-5}~g~cm^{-3}$.

In order to estimate the relevant cross sections
we should keep in mind that the proton is essentially scattered by the 
electrons,
whereas for the neutron, both proton and electron scattering processes should 
in principle be taken into account.  For these cases we have reproduced the 
estimates given in \cite{applegate}. 

Let us first consider the proton-electron
scattering. Here we use the expression from \cite{applegate}:
\begin{equation}
\sigma_{pe} = 2 \pi \int_{\theta_0}^{\pi} d \theta \; ( 1 - \cos \theta ) \;
\frac{2 \pi \alpha^2 m_e^2}{4 k^4 \sin^4(\frac{\theta}{2})} 
( 1 + \frac{k^2}{m_e^2} \cos^2(\frac{\theta}{2}) ) \;.
\label{eq:2.2}
\end{equation}
At $E \sim 1$ MeV, the Debye shielding of the proton by the electronic cloud
inhibits scattering below $\theta_0 \simeq 0.77 ^o$. Evaluating the integral
in eq. (\ref{eq:2.2})
we obtain  $\sigma_{pe} \simeq 1.5 \times 10^{-24}~cm^2$.  

The nucleon-nucleon cross
section at these energies is very well known given the large amount
of data available for the nucleon-nucleon phase shifts. 
Typically $\sigma_{np}$  is written as the sum of a singlet and a triplet 
contribution:
\begin{equation}
\sigma_{np} (E) = \frac{\pi a_s^2}{(a_s k)^2 + ( 1- 0.5 r_s a_s k^2)^2} +
\frac{3 \pi a_t^2}{(a_t k)^2 + (1 - 0.5 r_t a_t k^2)^2}\; ,
\label{eq:2.3}
\end{equation}
where the parameters for scattering lengths and scattering radii are obtained 
from fits to the data.
We have used  $a_s  =  -23.71$ fm;   $r_s =  2.73$ fm for the singlet component
and  $a_t = 5.432$ fm; $r_t = 1.749$ fm for the triplet component 
\cite{preston}.
This  yields the following value for the nucleon-nucleon cross section:
$\sigma_{np} \simeq 1.9 \times 10^{-24}~cm^2$. 

The neutrons interact with the electrons 
through their magnetic moment. At these energies ($E \sim 1$ MeV) the cross 
section
is given by \cite{applegate}:
\begin{equation}
\sigma_{ne} = 3 \pi  \: ( \frac{\alpha \:  K_{mag}}{m_n})^2 = 
8 \times 10^{-31} 
\mbox{ cm}^2\;,
\label{eq:2.4}
\end{equation}
where $ K_{mag} = - 1.91$ is the anomalous magnetic moment of the neutron
in nuclear magnetons.

The mean free path is defined as: $\lambda = \frac{m}{\rho \: \sigma}$ where
$m$ and $\rho$ are  the mass and density of the stopping 
particles and $\sigma$ is the relevant
stopping cross section.
Taking into account our estimates for the cross sections, 
we easily conclude that, in this epoch, both
neutrons and protons have mean free paths of the same order of magnitude:
$\lambda_p \sim \lambda_n \sim \lambda = 1~m $.

\begin{figure}[h]
\psfrag{alpha}{$\alpha_5$}
\psfrag{\lambda}{$\lambda_5$ (m)}
\centerline{
	\parbox[t]{0.5\textwidth}{
\centerline{\psfig{figure=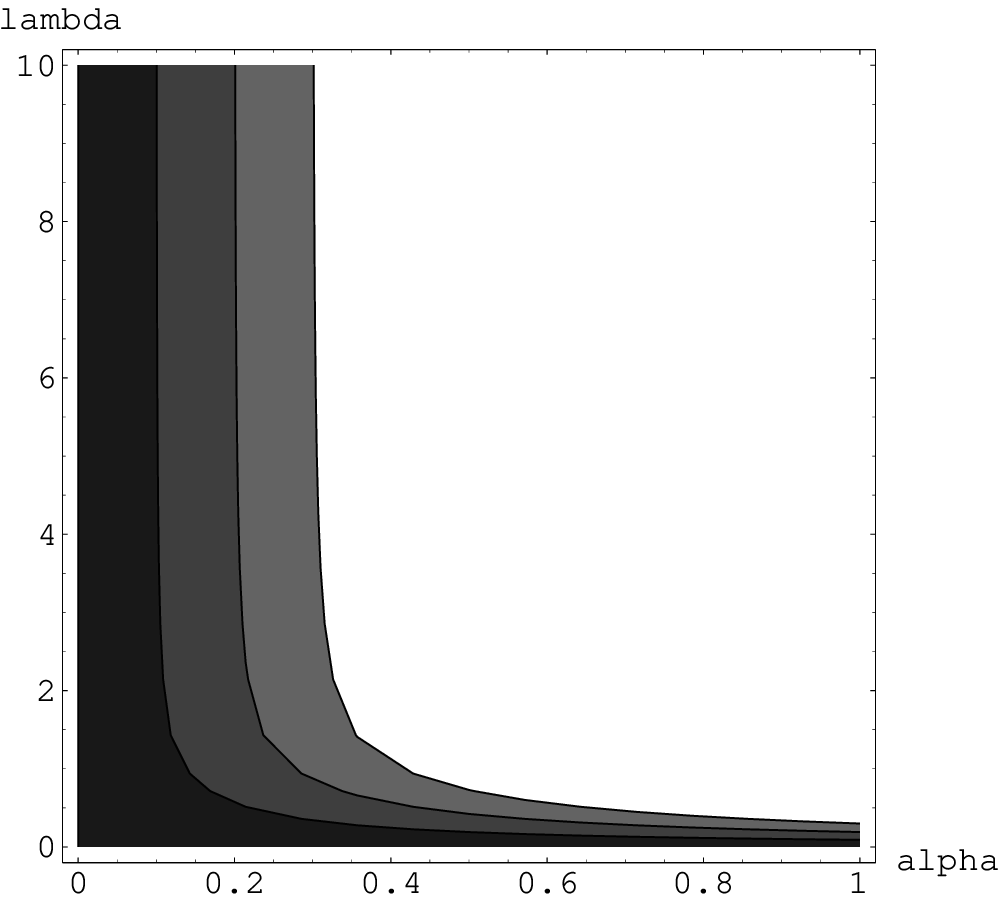,width=0.5\textwidth}}
	\caption{Limits imposed on the new
force parameters based on the nucleosysnthesis constraints for 
$\frac{\Delta G(r)}{G_{\infty}}$ within the 
ranges $[0,0.1]$,  $[0.1,0.2]$, and $[0.2,0.3]$
respectively.}	
\label{fig:1}}
}
\vspace{0.5cm}
\centerline{
	\parbox[t]{0.5\textwidth}{
\centerline{\psfig{figure=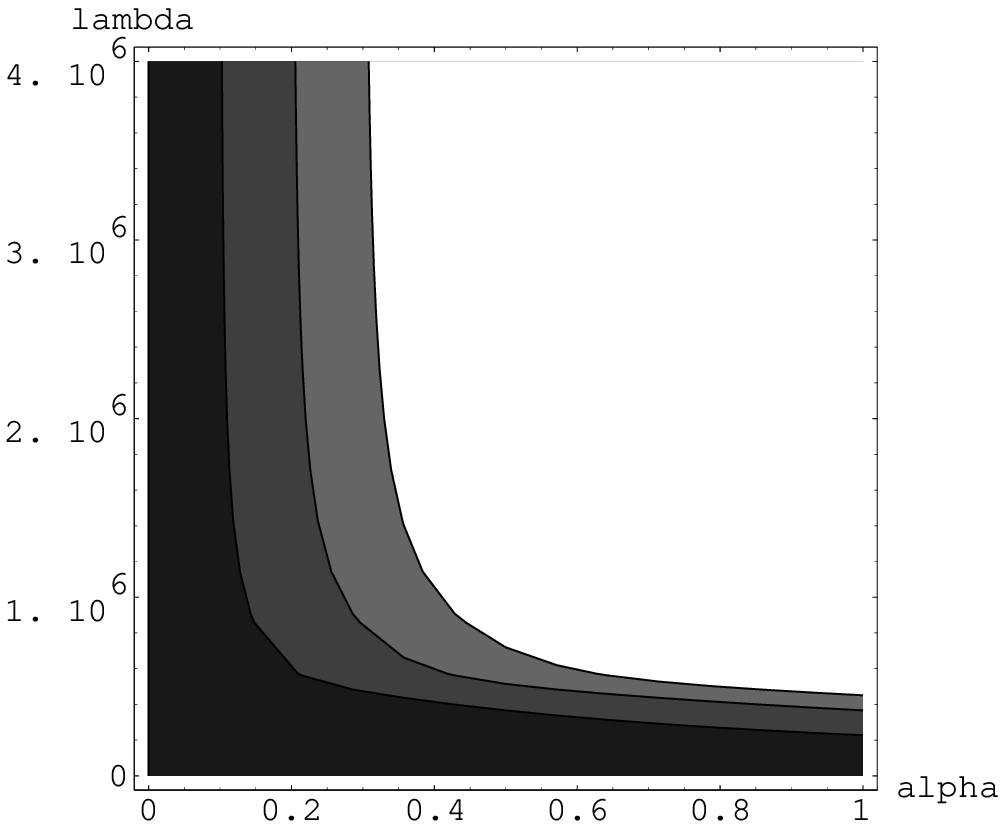,width=0.5\textwidth}}
	\caption{Limits imposed on the new
force parameters based on the nucleosysnthesis constraints and the 
inhomogeneous primordial nucleosysnthesis model for
$\frac{\Delta G(r)}{G_{\infty}}$ within the 
ranges $[0,0.1]$,  $[0.1,0.2]$, and $[0.2,0.3]$
respectively.}	
\label{fig:2}}
}
\end{figure}

In Figure (\ref{fig:1}) we present a contour plot of $\frac{\Delta 
G(r)}{G_{\infty}}$
as a function of the interaction parameters $(\alpha_5,\lambda_5)$ by taking
the typical distance between nucleons to be $r \sim  \lambda = 1~m$.  
The black, dark grey and light grey regions correspond to 
$\frac{\Delta G(r)}{G_{\infty}}$ within the 
ranges $[0,0.1]$,  $[0.1,0.2]$, and $[0.2,0.3]$
respectively.  One can immediately conclude from the contour shapes that,
whichever the range for the interaction, the maximum restriction that
can be extracted for the coupling constant $\alpha_5$  from
nucleosysnthesis corresponds exactly to the limit given for 
$\frac{\Delta G(r)}{G_{\infty}}$. It is worth underlining that 
the condition on $\alpha_5$ 
is valid for the primordial epoch (given that primordial
nucleosysnthesis was used). 
If $\alpha_5$ is assumed to be a running 
coupling constant, the limits here derived  for the
strength of this new  interaction are quite useful,  
given that laboratory experiments can never impose constraints on 
$\alpha_5^{prim}$.

Evidently $\alpha_5^{prim} \lsim 0.2$ is far above 
the typical values permitted by experiments 
($\alpha_5^{today} \lsim 10^{-4}$). 
In spite of that, even if the coupling constant is not running,
it is interesting to look at the
extremes where there are no experimental bounds. If this new interaction is
very short range $\lambda_5 < 10^{-3}~m$, 
primordial nucleosysnthesis does not constrain the
coupling constant at all. However  if this is a very long range 
interaction
($\lambda_5 > 10^{13}~m$),  we find that 
$\alpha_5^{prim} \lsim 0.2$ in order to satisfy eq. (\ref{eq:1.8}).

So far we have approximated the typical distance between
the interacting particles by the mean free paths 
$r \sim \lambda_n \sim \lambda_p$.
Alternatively,  a very trivial estimate for  the typical distance between 
nucleons could be extracted  directly from the total 
baryon density normalisation condition:
\begin{equation}
a=( \frac{1}{\rho_B N_A} )^{1/3} 
\label{eq:2.0}
\end{equation}
where $N_A$ is the Avogadro's number.
Evaluating eq.  (\ref{eq:2.0}) yields $a \sim 10^{-9}~m$. We find that now 
the limit $\alpha_5^{prim} \lsim 0.2$
holds even  if the interaction is short range $a < \lambda_5 < 10^{-3}~m$. 
Note again that there are no experimental limits on $\alpha_5^{today}$ 
in this region.

Ultimately  we shall consider the reasoning of the
previous section to obtain bounds
on the parameters of a new intermediate range interaction in the context of 
inhomogeneous nucleosynthesis scenario. 
In inhomogeneous Big Bang models one assumes
that there are non-linear perturbations of the baryon density producing
large proton deficient regions where the production of $^4$He is hindered 
\cite{applegate}.
The main advantage of these models is that they allow for a larger deuterium 
abundance,
which is one of the directions suggested by observational 
data \cite{steigman}.  

In this picture  most neutrons drift into neutron rich regions, and 
therefore we
consider that it is the  electron-neutron scattering process 
(rather than the proton scattering)
which determines the neutron mean free path.  
For the sake of our argument, we shall assume that the electrons remain
homogeneously distributed. Using the values of $\sigma_{ne}$, $\sigma_{pe}$ and
$\rho_e$ given in the previous section, we estimate a neutron mean free path
of  $\lambda_n^{Inhom} \sim 10^6~m$
whereas the proton mean free path remains essentially the same 
$\lambda_p \sim 1~m$.
If we now take the typical distance for interacting particles to be $r = 
\lambda_n^{Inhom}$
the restrictions imposed on the primordial value of the coupling constant are 
less severe.
This aspect is manifest in the contour plot shown in Figure (\ref{fig:2}).

We can then conclude that for $\lambda_5 \gsim  1~m$, the standard 
nucleosynthesis scenario is
more effective in constraining $\alpha_5$ then the inhomogeneous models.
%%%%%%%%%%%%%%%%%%%%%%%%%%%%%%%%%%%%%%%%%%%%%%%%%%

3. An independent estimate of the parameters of the new interaction 
can be obtained assuming, for instance, that the light boson carrier 
of the new 
interaction accounts for the main contribution of the dark matter 
in the galactic halo, as well as being responsible for the ionization
of interstellar hydrogen.
Even though this might seem highly speculative, the clumping of this 
bosonic field around and inside stars derives naturally 
from the coupling of this field to ordinary matter 
(cf. eq. (\ref{eq:3.1}) below). As discussed in Ref. \cite{ellis} this
implies that masses and couplings of ordinary particles depends on 
the features of the bosonic field interaction, meaning in turn, that the 
fundamental coupling constants are altered by the nearby density of matter.    
Estimates of the effects of this dependence on the cooling of neutron stars,
the neutrino burst in the supernova SN1987A and the period of the remnant
pulsar, imply bounds for $\alpha_5$ and $\lambda_5$ that are much less 
stringent than the ones emerging from E\"otv\"os-type experiments 
and from satellite measurements mentioned in the introduction. 
However, as we shall see,
interesting limits do arise if one assumes that the boson responsible 
for the new interaction decays into
photons and accounts for the dark matter contained in the halo of
our galaxy.

Denoting the light scalar field responsible for the new interaction by $\phi$,
we assume its coupling to nucleons and photons is of the following form:

\begin{equation}
{\cal L}_{int} = c_{n} {\phi \over \vev{\phi_5}} m_{n} N \bar N + c_{p} {\phi 
\over \vev{\phi_5}} F_{\mu \nu} F^{\mu \nu}~~,   
\label{eq:3.1}
\end{equation}
where $\vev{\phi_5}$ is a large scale associated to the new interaction. 
Of course, it is this scale that establishes the likelihood of 
creating this  field in particle physics accelerators.
This interaction yields, at tree level, a 
modification to the Newtonian potential such as that in eq. (\ref{eq:1.1}) 
where

\begin{equation}
\alpha_5 = {c_{n}^2 \over 4 \pi} \left({M_{p} \over \vev{\phi_5}}\right)^2~~, 
\label{eq:3.2}
\end{equation}  
and $M_P \equiv G_{\infty}^{-1/2}$ is the Planck mass. We see that 
the existing bounds on
$\alpha_5$ imply that $\vev{\phi_5} \sim M_P$. This agrees with what
had  already been mentioned in the introduction:
the new interaction must arise from physics close to the Planck scale. 
For simplicity we shall assume that $\vev{\phi_5} = M_P$. Following from
our previous conclusion that $\alpha_5^{prim} \lsim 0.2$ for 
$\lambda_5 \gsim 1~m$,  we obtain $c_{n} \lsim 1.58$.   
   
In order to extract further information on the new interaction, we 
demand that the scalar field, $\phi$, decays into photons that 
are energetic enough to account for the observed ionization 
of interstellar hydrogen, the temperature of Lyman-$\alpha$ clouds 
\cite{sciama1}, and the anomaly in the abundance of 
He I in the three high-redshift
Lyman-limit systems of the quasar HS $1700+6416$ \cite{sciama2}. This 
implies that $m_5 \gsim 27.2~eV$ and hence that 
$\lambda_5 \lsim 7 \times 10^{-9}~m$. This is consistent with the bounds 
on $\alpha_5$ obtained from supposing that the distance between nucleons during
nucleosynthesis is given by eq. (\ref{eq:2.0}).

Assuming the scalar particle is stable enough to account 
for all the dark matter
in the galactic halo, then the density of scalar particles 
is given by $\rho_5 = m_5 n_5 = \rho_{h}$, 
where $n_5$ is  the scalar particle number density and 
$\rho_{h} = 2 - 13 \times 10^{-25}~g~cm^{-3} = 0.1 - 0.7~GeV~cm^{-3}$ 
\cite{PDG} is the galactic halo density. This hypothesis yields that: 

\begin{equation}
3.67 \times 10^{6}~cm^{-3} < n_5 < 2.57 \times 10^{7}~cm^{-3}~~. 
\label{eq:3.3}
\end{equation}

Notice that if we had chosen to account for the observed cosmological 
energy density, we would have had to replace $\rho_h$ by 
$(0.2 - 0.3)\rho_{c}$, where 
$\rho_{c} = 1.88 \times 10^{-29}~h_{0}^2~g~cm^{-3} 
= 1.05 \times 10^{-5}~h_{0}^2~GeV~cm^{-3}$ is the critical density and 
$h_0$ is the Hubble parameter in units
of $100~km~s^{-1}~Mpc^{-1}$ (observationally $0.5 < h_0 < 0.8$).
Then the number density of scalar particles 
would be smaller than the estimate 
eq. (\ref{eq:3.3}) by about four orders of magnitude. 

To further verify our assumptions we compute the rate of decay of the scalar
particle. Since it can only decay into photons, its decay width can 
be obtained from the last term in eq. (\ref{eq:3.1}).  A straightforward
calculation reveals that:

\begin{equation}
\Gamma_5 = c_{n}^2 {m_{5}^3 \over \vev{\phi_5}^2}~~. 
\label{eq:3.4}
\end{equation}

\noindent
For $m_5 = 27.2~eV$, we have $t_5 \equiv \Gamma_5^{-1} 
= 3.0 \times 10^{36}~s >> t_U \approx H_{0}^{-1} =
h_0^{-1}~9.78~Gyr$. 
Thus the scalar particle responsible for the new interaction is  
comfortably stable to be a good dark matter candidate.

%%%%%%%%%%%%%%%%%%%%%%%%%%%%%%%%%%%%%%%%%%

4. In this paper we have discussed bounds on the parameters of a new 
interaction that couples with baryon number, arising from
primordial nucleosynthesis. We have considered the standard nucleosynthesis 
scenario
and shown that $\alpha_5 \lsim 0.2$ for $\lambda_5 \gsim 1~m$ (see Figure 1).
The inhomogeneous nucleosynthesis scenario of Ref. \cite{applegate} allows
one to conclude the same for $\lambda_5 \gsim 5 \times 10^{5}~m$ (Figure 2).
These limits do not require the assumption of a running coupling 
constant and even though consistent, they are much less stringent than 
the ones obtained from laboratory experiments and satellite tests. 
As discussed in the text,
they may be relevant for a new interaction whose coupling constant is
scale-dependent and is not asymptotically free. 
We have also derived bounds on the clumpiness of the scalar particle
assuming it is the main contributor to the dark matter in the galactic
halo, that is $3.67 \times 10^{6}~cm^{-3} < n_5 < 2.57 \times 10^{7}~cm^{-3}$ 
for $\lambda_5 \lsim 7 \times 10^{-9}~m$. This corresponds to  
$m_5 \gsim 27.2~eV$ implying that the scalar particle decay can account for the
ionization of hydrogen observed in the interstellar medium and the temperature
of Lyman-$\alpha$ clouds.

\vspace{0.5cm}

{\large\bf Acknowledgements} 

\vspace{0.5cm}
\noindent
One of us (F.N.) gratefully acknowledges the 
support of the JNICT Fellowship BIC 1481.

\newpage

\end{document}